\documentclass[twocolumn,aps,showpacs,superscriptaddress]{revtex4-1}
\usepackage{amsmath}
\usepackage{graphicx}
\usepackage{slashed}
\RequirePackage[pagewise,mathlines]{lineno}

\begin{document}
\title{Initial transverse-momentum broadening of Breit-Wheeler process in relativistic heavy-ion collisions}%
\author{W. Zha}\affiliation{University of Science and Technology of China, Hefei, China}
\author{J.D. Brandenburg}\affiliation{Shandong University, Qingdao, China}\affiliation{Brookhaven National Laboratory, New York, USA}\affiliation{Rice University, Houston, Texas, USA}
\author{Z. Tang}\affiliation{University of Science and Technology of China, Hefei, China}
\author{Z. Xu}\email{xzb@bnl.gov}\affiliation{Shandong University, Qingdao, China}\affiliation{Brookhaven National Laboratory, New York, USA}
\date{\today}%
\begin{abstract}
We calculate the cross section and transverse-momentum ($P_{\bot}$) distribution of the Breit-Wheeler process in relativistic heavy-ion collisions and their dependence on collision impact parameter ($b$). To accomplish this, the Equivalent Photon Approximation (EPA) was generalized in a more differential way compared to the approach traditionally used for inclusive collisions. In addition, a lowest-order QED calculation with straightline assumption was performed as a standard baseline for comparison. The cross section as a function of $b$ is consistent with previous calculations using the equivalent one-photon distribution function. Most importantly, the $P_{\bot}$ shape from this model is strongly dependent on impact parameter and can quantitatively explain the $P_{\bot}$ broadening observed recently by RHIC and LHC experiments. This broadening effect from the initial QED field strength should be considered in studying possible trapped magnetic field and multiple scattering in a Quark-Gluon Plasma (QGP). The impact-parameter sensitive observable also provides a controllable tool for studying extreme electromagnetic fields.

\end{abstract}
\maketitle
In 1934, Breit and Wheeler studied the process of collision of two light quanta~\cite{Breit-wheeler1934zz} to create electron and positron pairs. The Weizsaeker-William ($WW$) photon flux generated by highly charged heavy ions in a moving frame of reference was proposed as the only viable approach to achieve photon-photon collisions. In recent years, high-energy particle accelerators~\cite{SLACPhysRevLett.79.1626} with high luminosity and high-power lasers~\cite{Piazza2012RevModPhys.84.1177,Pike2014NaturePhotonics} with fast pulses are able to generate extreme electromagnetic fields and realize photon-photon collisions. In the collider mode of relativistic heavy-ion collisions, the intense electromagnetic fields are equivalent to a flux of high-energy $WW$ photons~\cite{BAUR20071,KRAUSS1997503} in the laboratory frame, and the production of dileptons can be represented as the product of two-photon collisions $\gamma + \gamma \rightarrow l^{+} + l^{-} $. The magnetic field generated by these passing nuclei can reach $10^{15}$~Tesla~\cite{BFieldKharzeev:2007jp} and has become an important experimental and theoretical element in the study of new QCD phenomena~\cite{ReportKharzeev:2015znc}.

From external classical field~\cite{PhysRevC.47.2308} and EPA approximations, photons are generated by the QED field with momentum predominantly along the beam direction and with transverse momentum at the scale of $\omega/\gamma$ where $\omega$ is the photon energy and $\gamma$ is the Lorentz factor of the projectile and target nuclei. Higher-order contributions are suppressed by orders of $1/\gamma^2$ and are negligible for the case when the $P_\bot$ of the photon is of $\omega/\gamma$ or less. Therefore, the associated photons have a small $P_{\bot}$ and the leptons subsequently produced by these QED fields in the EPA have the distinctive signature of being nearly back-to-back in azimuth with small total transverse momenta ($P_\bot\simeq\omega/\gamma$). The photon-induced scattering processes have been extensively studied in ultra-peripheral collisions (UPCs)~\cite{PhysRevC.70.031902,DYNDAL2017281,Abbas2013,2017489,BAUR1990786,Klein:2016yzr,PhysRevC.80.044902,Hencken:1995me,0954-3899-15-3-001,PhysRevC.47.2308,Alscher:1996mja,ATLASNatureAaboud:2017bwk}, for which the impact parameter ($b$) is larger than twice the radius ($R_{A}$) of the nucleus to ensure that no hadronic interaction occurs between the colliding nuclei. Conventionally, the photon-induced interactions are expected only to be applicable in UPCs. However, in the last few years, such photo-processes have also been observed in hadronic heavy-ion collisions (HHICs) for photonuclear production~\cite{LOW_ALICE,1742-6596-779-1-012039} and photon-photon collisions~\cite{syang2016thesis,PhysRevLett.121.132301,PhysRevLett.121.212301}, and theoretical progress~\cite{PhysRevC.93.044912,PhysRevC.97.044910,ZHA2018182,PhysRevC.97.054903} has been made to describe such processes. 

Furthermore, the STAR collaboration at RHIC~\cite{PhysRevLett.121.132301} and the ATLAS collaboration at the LHC~\cite{PhysRevLett.121.212301} have found a significant $P_{\bot}$ broadening effect for the lepton pairs from photon-photon collisions in HHICs in comparison to those in UPCs and to model calculations. The STAR Collaboration characterized the broadening by measuring the $P_{\bot}^{2}$ and the invariant mass spectra of lepton pairs in Au+Au and U+U collisions with respect to the EPA calculations. The ATLAS Collaboration quantified the effect via the acoplanarity of lepton pairs in different centrality HHIC in contrast to the same measurements in UPCs. Theoretical models are used to describe the broadening qualitatively by introducing the effect of a magnetic field trapped in an electrically conducting QGP or alternatively by the electromagnetic (EM) scattering of leptons in the hot and dense medium. These descriptions of the broadening effect assume that there is no impact-parameter dependence of the $P_{\perp}$ distribution for the lepton pair from the initial photon-photon collision. In this letter, we examine the Breit-Wheeler process as a function of impact parameter in heavy-ion collisions based on the external classical field approach~\cite{PhysRevC.47.2308} and go through the generalized EPA (gEPA1) assumption toward a formula, which shows strong impact-parameter dependence of the cross section and $P_\bot$ spectra. We have pointed out possible issue with this EPA approach, proposed an empirical remedy with a modified formula (gEPA2) and also carried out a full QED calculation based on Ref.~\cite{PhysRevA.51.1874,PhysRevA.55.396}. These contributions provide a baseline for the extraction of possible medium effect. 

Following the procedure of external classical field approach in~\cite{PhysRevC.47.2308}, we start from the electromagnetic potentials of the two colliding nuclei in the Lorentz gauge:
    \begin{equation}
    \label{equation1}
    \begin{split}
    &A_{1}^{\mu}(k_{1},b)= -2 \pi (Z_{1} e) e^{ik_{1}^{\tau}b_{\tau}} \delta(k_{1}^{\nu}u_{1\nu}) \frac{F_{1}(-k_{1}^{\rho}k_{1\rho})}{k_{1}^{\sigma}k_{1\sigma}} u_{1}^{\mu},\\
     &A_{2}^{\mu}(k_{2},0)= -2 \pi (Z_{2} e) e^{ik_{2}^{\tau}b_{\tau}} \delta(k_{2}^{\nu}u_{2\nu}) \frac{F_{2}(-k_{2}^{\rho}k_{2\rho})}{k_{2}^{\sigma}k_{2\sigma}} u_{2}^{\mu}.\\
    \end{split}
    \end{equation}
Here $b$ is the impact parameter, which characterizes the separation between the two nuclei. The $\delta$ function ensures that the nuclei travel in straight line motion with a constant velocity. In this approach, the deflections from the straight line motion due to collisions are neglected. The velocities are taken in the center-of-mass frame with $u_{1,2} = \gamma(1,0,0,\pm v)$, where $\gamma$ is the Lorentz contraction factor. The form factor $F(k^2)$ is the nuclear electromagnetic form factor obtained from the Fourier transformation of the charge density of the nucleus. The amplitude for the lepton pair production from the electromagnetic fields in lowest order can be given by the S-matrix element, which leads to Eq.~30 in Ref.~\cite{PhysRevC.47.2308}:

   \begin{equation}
    \label{equation12}
    \begin{split}
    &\sigma = 16\frac{Z^{4}e^{4}}{(4\pi)^{2}}\int d^{2}b \int \frac{dw_{1}}{w_{1}} \frac{dw_{2}}{w_{2}} \frac{d^{2}k_{1\bot}}{(2\pi)^{2}} \frac{d^{2}k_{2\bot}}{(2\pi)^{2}} \frac{d^{2}q_{\bot}}{(2\pi)^{2}}\\
     &\times \frac{F(-k_{1}^{2})}{k_{1}^{2}} \frac{F(-k_{2}^{2})}{k_{2}^{2}} \frac{F^{*}(-{k_{1}^{\prime}}^{2})}{{k_{1}^{\prime}}^{2}} \frac{F^{*}(-{k_{2}^{\prime}}^{2})}{{k_{2}^{\prime}}^{2}} e^{-i\vec{b} \cdot \vec{q}_{\bot}} \\
     & \times \big[(\vec{k}_{1\bot} \cdot \vec{k}_{2\bot})(\vec{k}^{\prime}_{1\bot} \cdot \vec{k}^{\prime}_{2\bot}) \sigma_{s}(w_{1},w_{2})\\
     & +(\vec{k}_{1\bot} \times \vec{k}_{2\bot})(\vec{k}^{\prime}_{1\bot} \times \vec{k}^{\prime}_{2\bot})\sigma_{ps}(w_{1},w_{2})\big] 
    \end{split}
    \end{equation}
where the four momenta of photons are
   \begin{equation}
    \label{equation4}
    \begin{split}
    k_{1} &= (w_{1}, k_{1\bot},\frac{w_{1}}{v}), k_{2} = (w_{2}, P_{\bot} - k_{1\bot},\frac{w_{2}}{v})\\
     &w_{1} =\frac{1}{2}(P_{0}+vP_{z}), w_{2} = \frac{1}{2}(P_{0}-vP_{z}) \\
     &k_{2\bot} = P_{\bot} - k_{1\bot}, q_{\bot} = k_{1\bot} -k_{1\bot}^{\prime} \\
     &k_{1}^{\prime} =(w_{1}, k_{1\bot} - q_{\bot},w_{1}/v)\\ &k_{2}^{\prime} =(w_{2}, k_{2\bot} + q_{\bot},w_{2}/v)\\
    \end{split}
    \end{equation}

\begin{figure*}[htbp]
\includegraphics[keepaspectratio,width=0.99\textwidth]{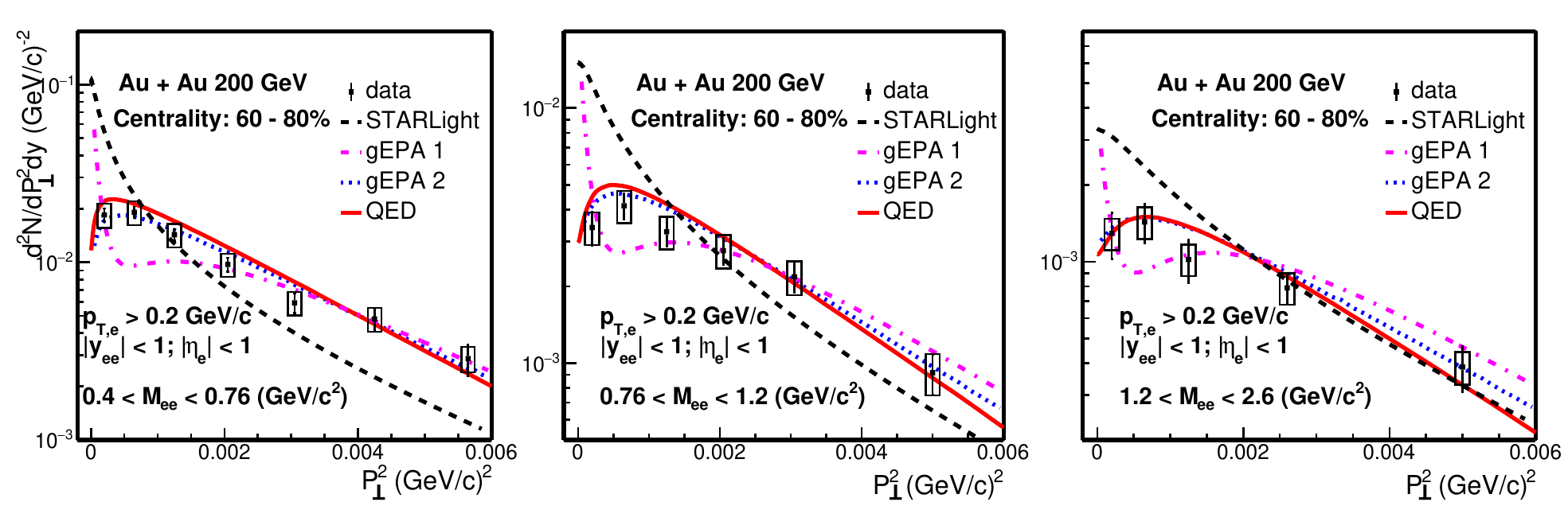}
\caption{The $P_{\bot}^{2}$ distributions of electron-positron pair production within the STAR acceptance for the mass regions $0.4-0.76$ (left panel), $0.76-1.2$ (middle panel), and $1.2-2.6$ GeV$/c^{2}$ (right panel) in $60-80\%$ Au+Au collisions at $\sqrt{s_{\rm{NN}}} =$ 200 GeV. The STAR measurements~\cite{PhysRevLett.121.132301} and calculations from gEPA1, gEPA2, QED and STARLight~\cite{Klein:2016yzr} are also plotted for comparison. See text for details of the models.}
\label{figure1}
\end{figure*}

The elementary scalar ($\sigma_{s}$) and pseudoscalar ($\sigma_{ps}$) cross-sections can be given by the following formula
     \begin{equation}
  \label{equation13}
  \begin{aligned}
  & \sigma_{s} = \frac{4\pi \alpha^{2}}{s} \bigg[\left(2+\frac{8m^{2}}{s} - \frac{24m^{4}}{s^{2}}\right)\text{ln}\left(\frac{\sqrt{s}+\sqrt{s-4m^{2}}}{2m}\right)
  \\
  & -\sqrt{1-\frac{4m^{2}}{s}}\left(1+\frac{6m^{2}}{s}\right)\bigg]\\
    & \sigma_{ps} = \frac{4\pi \alpha^{2}}{s} \bigg[\left(2+\frac{8m^{2}}{s} - \frac{8m^{4}}{s^{2}}\right)\text{ln}\left(\frac{\sqrt{s}+\sqrt{s-4m^{2}}}{2m}\right)
  \\
  & -\sqrt{1-\frac{4m^{2}}{s}}\left(1+\frac{2m^{2}}{s}\right)\bigg],\\
  \end{aligned}
  \end{equation}
where $m$ is the lepton mass and $s=4w_{1}w_{2}$. To arrive at Eq.~\ref{equation12} from Eq.~\ref{equation1}, it has been shown in Ref.~\cite{PhysRevC.47.2308} that the photon flux is decisively longitudinal with 
$k_{\bot}\simeq\omega/\gamma$ and higher-order terms are of the order of $1/\gamma^2$ down from the first term presented in Eq.~\ref{equation12}. This is also the only approximation from the external classic field resulting in the equivalent photon flux. 
For differential cross-section as a function of a given variable (for example, $P_{\perp}$ or $b$), the integration is performed 
over all other phase spaces except the given variable ($P_{\perp}$ or $b$). It has been pointed out that the equations are used for 
total cross-section without kinematic cuts. The final-state dilepton pair kinematics were implemented and selected according to 
experimental acceptance using the readily available code 
from STARLight~\cite{Klein:2016yzr} with the Breit-Wheeler formalism for the differential cross-section in $\gamma\gamma\rightarrow l^{+}l^{-}$.  Unless otherwise specified, the cross-sections of $\sigma_{s}$ and $\sigma_{ps}$ are specific cross-sections within the detector acceptance. 
The results from this model are shown in Fig.~\ref{figure1} and are labelled as generalized EPA 1 (gEPA1). 

\renewcommand{\floatpagefraction}{0.75}

Through Eq.~\ref{equation12} and~\ref{equation13}, we are able to study the impact parameter dependence of the differential cross-section on the $P_{\bot}$ spectra of lepton pairs. To simplify the numerical calculation, we assume that the charges in the target and projectile nuclei are distributed according to the Woods-Saxon distribution without any fluctuations or point-like structure. The charge density for a symmetrical nucleus $A$ is given by the Woods-Saxon distribution:
   \begin{equation}
   \rho_{A}(r)=\frac{\rho^{0}}{1+\exp[(r-R_{\rm{WS}})/d]}
   \label{equation17}
   \end{equation}
   where the radius $R_{\rm{WS}}$ (Au: 6.38 fm, Pb: 6.62fm) and skin
depth $d$ (Au: 0.535 fm, Pb: 0.546 fm) are based on fits to electron
scattering data~\cite{0031-9112-29-7-028}, and $\rho^{0}$ is the
normalization factor. The collision geometry (centrality) is determined by the Glauber model,
in which the p+p inelastic cross sections employed for RHIC and LHC are
42 mb and 70 mb at their corresponding collision energies, respectively. Fig.~\ref{figure1} shows the numerical calculations of the $P_{\perp}^{2}$ distributions of electron-positron pair production for different mass regions in 60-80$\%$ Au+Au collisions at $\sqrt{s_{\rm{NN}}} =$ 200 GeV. The results are filtered with the STAR acceptance ($p_{T,e} > 0.2 $ GeV/c, $|\eta| <1$, and $|y_{ee}|<1$) to allow direct comparison with the experimental measurements. The STAR measurements~\cite{PhysRevLett.121.132301} and calculations from the STARLight EPA approach are also plotted for comparison. Both the shape and the magnitude of the experimental measurements can be described reasonably well by these calculations. However, there is a notable wiggle at low $P_\perp$, which disagrees with the data and appears to be unphysical. 

 \renewcommand{\floatpagefraction}{0.60}

Usually, an integration over $b$ of Eq.~\ref{equation12} is performed resulting in the famous equivalent photon cross-section for two-photon collisions as shown in Eq.~32 of Ref.~\cite{PhysRevC.47.2308}, as in STARLight and other numerical calculations~\cite{Klein:2016yzr,Klein:2018fmp}. The integration of Eq.~\ref{equation12} over the impact parameter $b$ leads to a $\delta$ function in the transverse momentum $q_{\bot}$ and the subsequent result reads:
   \begin{equation}
    \label{equation14}
    \begin{split}
    \sigma = 16\frac{Z^{4}e^{4}}{(4\pi)^{2}} \int &\frac{dw_{1}}{w_{1}} \frac{dw_{2}}{w_{2}} \frac{d^{2}k_{1\bot}}{(2\pi)^{2}} \frac{d^{2}k_{2\bot}}{(2\pi)^{2}} \bigg|\frac{F(-k_{1}^{2})}{k_{1}^{2}}\bigg|^{2}\\
     &\times \bigg|\frac{F(-k_{2}^{2})}{k_{2}^{2}}\bigg|^{2} k_{1\bot}^{2} k_{2\bot}^{2} \sigma(w_{1},w_{2})
    \end{split}
    \end{equation}
where $\sigma(w_{1},w_{2})$ is the cross-section averaged over the scalar and pseudoscalar polarization. This is exactly the EPA expression commonly used in the literature and used in comparison to recent experiments~\cite{KRAUSS1997503}. 
The spectral shape~\cite{Klein:2016yzr,Klein:2018fmp}, which is insensitive to the collision centrality, is the result of integrating over the whole impact parameter space as shown in Eq.~31 to Eq.~32~\cite{PhysRevC.47.2308} and subsequently inserting an impact-parameter dependent photon flux $\sigma(w_{1},w_{2}, b)$, as shown in Eq.~36 to 43 in Ref.~\cite{PhysRevC.47.2308}. 

We have also performed a QED calculation at leading-order based on Ref~\cite{PhysRevA.51.1874,PhysRevA.55.396} and extended its original calculation to all impact parameters as a function of the transverse momentum of the produced pair. 
The lowest-order two-photon interaction is a second-order process with two Feynman diagrams contributing, as shown in Fig.~2 of Ref.~\cite{PhysRevA.51.1874,PhysRevA.55.396}. Similarly, the straight-line approximation for the incoming projectile and target nuclei is applied as in the case of all EPA calculations. 
Otherwise, a full QED calculation of the differential cross-section with two photons colliding to create two leptons has been calculated. 
Following the derivation of Ref.~\cite{PhysRevA.51.1874,PhysRevA.55.396}, the cross section for pair production of leptons is given by
   \begin{equation}
    \label{equation1_new}
\sigma = \int d^{2}b \frac{d^{6}P(\vec{b})}{d^{3}p_{+}d^{3}p_{-}} = \int d^{2}q \frac{d^{6}P(\vec{q})}{d^{3}p_{+}d^{3}p_{-}} \int d^{2}b e^{i {\vec{q}} \cdot  {\vec{b}}},
    \end{equation}
and the differential probability $\frac{d^{6}P(\vec{q})}{d^{3}p_{+}d^{3}p_{-}}$ in QED at the lowest order is 
\begin{equation}
\label{equation2_new}
\begin{split}
&\frac{d^{6}P(\vec{q})}{d^{3}p_{+}d^{3}p_{-}} = (Z\alpha)^{4}
\frac{4}{\beta^{2}} \frac{1}{(2\pi)^{6}2\epsilon_{+}2\epsilon_{-}} \int d^{2}q_{1}\\
& F(N_{0})F(N_{1})F(N_{3})F(N_{4})[N_{0}N_{1}N_{3}N_{4}]^{-1} \\
& \times {\rm{Tr}}\{(\slashed{p}_{-}+m)[N_{2D}^{-1}\slashed{u}_{1} (\slashed{p}_{-} - \slashed{q}_{1} + m)\slashed{u}_{2} + \\
& N_{2X}^{-1}\slashed{u}_{2}(\slashed{q}_{1} - \slashed{p}_{+} +m)\slashed{u}_{1}] (\slashed{p}_{+}-m)[N_{5D}^{-1}\slashed{u}_{2}\\
& (\slashed{p}_{-} - \slashed{q}_{1} - \slashed{q} + m)\slashed{u}_{1} + N_{5X}^{-1} \slashed{u}_{1}(\slashed{q}_{1} + \slashed{q} - \slashed{p}_{+} \\
& + m)\slashed{u}_{2}] \},
\end{split}
    \end{equation}
with
  \begin{equation}
    \label{equation3_new}
    \begin{split}
 & N_{0} = -q_{1}^{2},  N_{1} = -[q_{1} - (p_{+}+p_{-})]^{2},\\
 & N_{3} = -(q_{1}+q)^{2}, N_{4} = -[q+(q_{1} - p_{+} - p_{-})]^{2}, \\
 & N_{2D} = -(q_{1} - p_{-})^{2} + m^{2},\\
  & N_{2X} = -(q_{1} - p_{+})^{2} + m^{2}, \\
 &N_{5D} = -(q_{1} + q - p_{-})^{2} + m^{2},\\
  & N_{5X} = -(q_{1} + q  - p_{+})^{2} + m^{2},
\end{split}
    \end{equation}
where $p_{+}$ and $p_{-}$ are the momenta of the created leptons, the longitudinal components of $q_{1}$ are given by $q_{10} = \frac{1}{2}[(\epsilon_{+} + \epsilon_{-}) + \beta(p_{+z}+p_{-z})]$, $q_{1z} = q_{10}/ \beta$, $\epsilon_{+}$ and $\epsilon_{-}$ are the energies of the produced leptons, and $m$ is the mass of lepton. In the calculation of $P(\vec{q})$, the traces and matrices have been handled by the Mathematica package FeynCalc~\cite{SHTABOVENKO2016432}. The multi-dimensional integration is performed with the Monte Carlo (MC) integration routine VEGAS~\cite{PETERLEPAGE1978192}.

The gEPA1 and QED calculations are shown in Fig.~\ref{figure1} as dash-dotted and solid lines, respectively, together with experimental data points and the STARLight calculations. 
It is clear that there is a difference between the gEPA1 and the QED calculations. 
The most striking difference is in the $P_{\perp}$ spectral shape.  
The QED curves describe the spectra quite well with a smooth distribution of the cross-section increasing from high to low $P_{\perp}$, but with a smooth turn-over at very low $P_{\perp}$. 
We stipulated that the two Feynman diagrams in the leading order have interference terms which is missing from the generalized EPA (gEPA1) in the connection between initial virtual photon fields and the Breit-Wheeler cross-section. 
We tried an additional phase term for the differential cross-section leading Eq.~\ref{equation12} to become: 

   \begin{equation}
    \label{equation16}
    \begin{split}
    &{\frac{d\sigma}{P_{\bot}dP_{\bot}}} = 16\frac{Z^{4}e^{4}}{(4\pi)^{2}}\int d^{2}b \int \frac{dw_{1}}{w_{1}} \frac{dw_{2}}{w_{2}} \frac{d^{2}k_{1\bot}}{(2\pi)^{2}} \frac{d^{2}q_{\bot}}{(2\pi)^{2}}\\
     &\times \frac{F(-k_{1}^{2})}{k_{1}^{2}} \frac{F(-k_{2}^{2})}{k_{2}^{2}} \frac{F^{*}(-{k_{1}^{\prime}}^{2})}{{k_{1}^{\prime}}^{2}} \frac{F^{*}(-{k_{2}^{\prime}}^{2})}{{k_{2}^{\prime}}^{2}} \\
     &\times e^{-i\vec{b} \cdot \vec{q}_{\bot}+i\alpha_1+i\alpha_2} \\
     & \times \big[(\vec{k}_{1\bot} \cdot \vec{k}_{2\bot})(\vec{k}^{\prime}_{1\bot} \cdot \vec{k}^{\prime}_{2\bot}) \sigma_{s}(w_{1},w_{2})\\
     & +(\vec{k}_{1\bot} \times \vec{k}_{2\bot})(\vec{k}^{\prime}_{1\bot} \times \vec{k}^{\prime}_{2\bot})\sigma_{ps}(w_{1},w_{2})\big] 
    \end{split}
    \end{equation}
where $\alpha_1$ is the angle between $\vec{k}_{1\bot}$ and $\vec{k}^{\prime}_{1\bot}$ while 
$\alpha_2$ is the angle between $\vec{k}_{2\bot}$ and $\vec{k}^{\prime}_{2\bot}$. The results are shown in Fig.~\ref{figure1} as gEPA2. 
The gEPA2 result is strikingly similar to the QED calculation in all centralities and beam energies. 
Although there is no vigorous physics motivation for the additional phase shift, it is nevertheless valuable to be taken as an effective way of empirical parameterization of the data. 
It certainly will be interesting to see if such a description holds with new precision data from UPCs at RHIC and LHC. 
We note that there is a small difference between the QED and the gEPA2 results evident in Fig.~\ref{figure1}. Specifically, the QED calculation has a systematically lower $\sqrt{\langle P_{\bot}^{2} \rangle }$ than the STAR data and the gEPA2 calculation.
Whether this is a result from final-state interaction or from missing ingredients in the QED calculation in the initial stage requires further theoretical and experimental exploration.  
One possibility is to compare the QED calculation to experimental data in UPC to validate the calculations. 

\begin{figure}[htbp]
\includegraphics[keepaspectratio,width=0.45\textwidth]{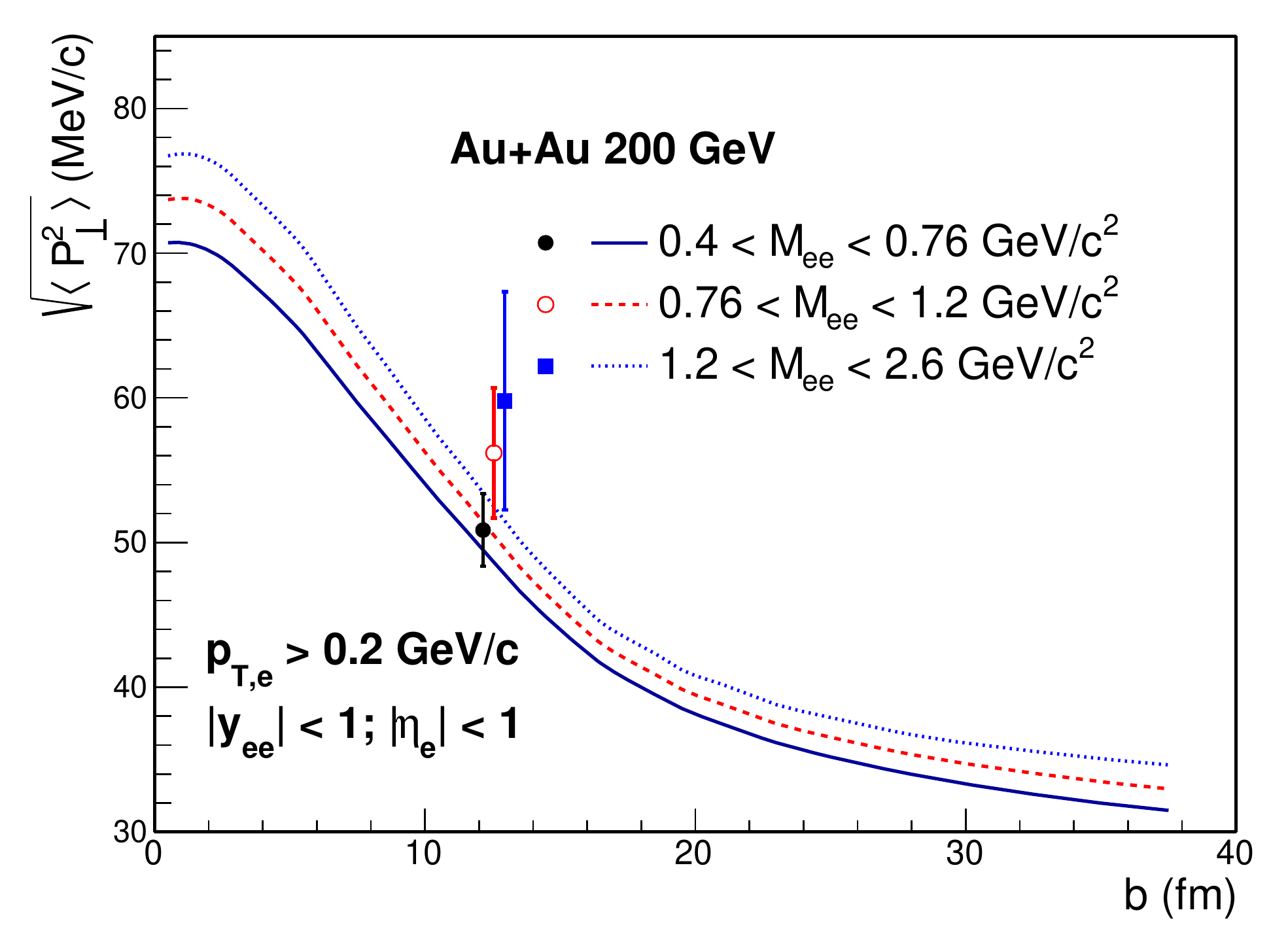}
\caption{The $\sqrt{\langle P_{\bot}^{2} \rangle}$ of electron-positron pairs within the STAR acceptance as a function of the impact parameter $b$ for different mass regions in Au+Au collisions at $
\sqrt{s_{\rm{NN}}} =$ 200 GeV. The STAR measurements~\cite{PhysRevLett.121.132301} are also plotted for comparison. The data points are slightly shifted for clarity.}
\label{figure2}
\end{figure}

As demonstrated in Eq.~\ref{equation12},~\ref{equation1_new} and \ref{equation16}, the term $e^{i\vec{b} \cdot \vec{q}_{\bot}}$ gives rise to impact parameter dependence in the integration over the transverse momentum $q_{\bot}$. 
Figure~\ref{figure2} shows the numerical QED calculations of the broadening variable, $\sqrt{\langle P_{\bot}^{2} \rangle }$, for electron-positron pairs within STAR acceptance as a function of impact parameter $b$ for different mass regions in Au+Au collisions at $\sqrt{s_{\rm{NN}}} =$ 200 GeV.
As expected, the broadening increases with deceasing impact parameter and reaches a maximum value approximately twice that of the peripheral collisions. 
In addition, the broadening depends on the invariant mass of the lepton pair. 
The larger the invariant mass, the more significant the broadening. 
A possible explanation is that the pairs with larger invariant mass are generated predominantly in the vicinity of the stronger electromagnetic field, which, in turn, create higher transverse momentum. 
In this figure, the STAR measurements~\cite{PhysRevLett.121.132301} are also plotted for comparison and show qualitative agreement but with systematically higher values than the QED calculations. 
We summarize as follows: 
1. The $\sqrt{\langle P_{\bot}^{2} \rangle }$ from data requires an additional $\simeq40$ MeV/$c$ broadening compared to the STARLight result~\cite{PhysRevLett.121.132301}; 
2. The QED calculation in 60-80\% requires $\simeq30$ MeV/$c$ additional broadening compared to its UPC result; 
3. The $\sqrt{\langle P_{\bot}^{2} \rangle }$ from STAR data in 60-80\% central collisions is about 15 MeV/$c$ broader than the QED calculation at the same centrality. 
This means that the final-state broadening effect, if any, is about a factor of 3 smaller than the original estimate~\cite{PhysRevLett.121.132301}. 
We noted that Ref.~\cite{Hencken:2004td} has used similar approach with a minimum cutoff on impact parameter to reproduce an earlier STAR result in ultra-peripheral collisions under the condition of mutual Coulomb excitation~\cite{PhysRevC.70.031902}.

\renewcommand{\floatpagefraction}{0.75}

\begin{figure*}[htbp]
\includegraphics[keepaspectratio,width=0.99\textwidth]{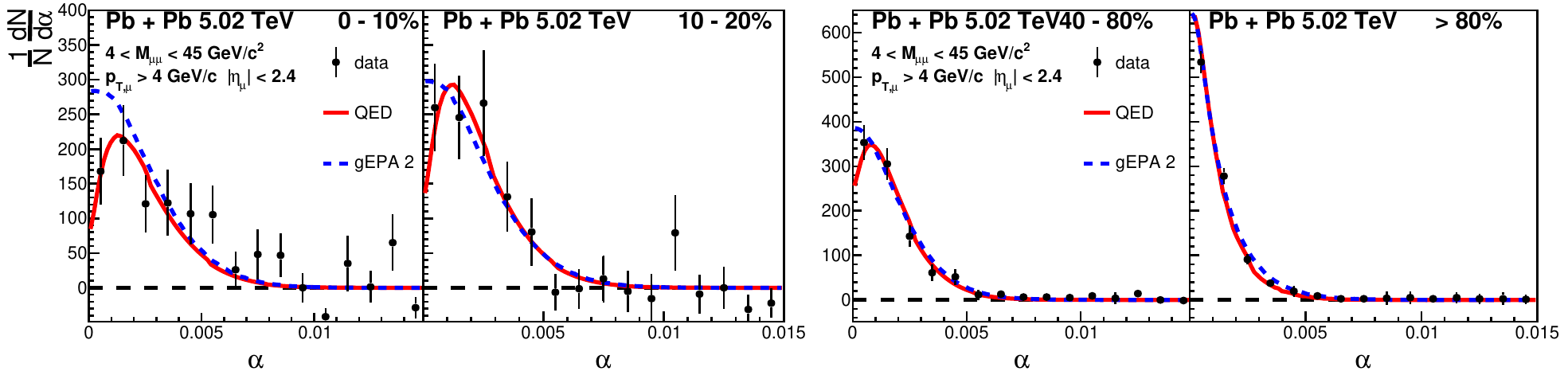}
\caption{The distributions of the broadening variable, $\alpha$, from the generalized EPA approach (gEPA2, dash blue lines) and QED (solid red line) for muon pairs in Pb+Pb collisions at $\sqrt{s_{\rm_{NN}}} = $ 5.02 TeV for different centrality classes. The results are filtered with the fiducial cuts described in the text and normalized to unity to facilitate a direct comparison with experimental data. The measurements from ATLAS~\cite{PhysRevLett.121.212301} are also plotted for comparison.}
\label{figure3}
\end{figure*}

 The STAR Collaboration performed this measurement with electron-positron pairs in the mass range of [0.4,2.6] GeV/$c^2$. The $\omega/\gamma$ is in the range of [4,26] MeV and is at the same order of the measured $P_\bot$ range as a required condition of EPA~\cite{PhysRevC.70.031902}. 
 The measurement of a smooth structure in the invariant mass distribution with the absence of vector mesons ($\rho, \omega$ and $\phi$) is convincing evidence of pure Breit-Wheeler process~\cite{syang2016thesis}. 
 The excellent electron identification at low momentum along with large coverage and low material budget along the trajectories of produced particles at mid-rapidity makes this difficult measurement possible at STAR. 
 The ATLAS Collaboration takes advantage of better angular measurement (than momentum measurement) of high-momentum muons, and characterizes the broadening of muon pairs with acoplanarity correlations in the mass range of $ 4 < M_{\mu\mu} < 45$ GeV/$c^{2}$ from peripheral to central collisions~\cite{PhysRevLett.121.212301}. 
 The observed broadening by the pair acoplanarity, $\alpha$, is defined as 

\begin{equation}
\label{equation15}
\alpha = 1 - \frac{|\phi^{+} - \phi^{-}|}{\pi}
\end{equation}

where $\phi^{\pm}$ represent the azimuthal angles of the two individual muons. 
The acoplanarity was use to avoid the detector-induced distortions from poor momentum resolution. 
It should be noted that $\pi\alpha\simeq P_{\bot}/M_{ll}$ in a detector setup where the sagitta of a particle trajectory is much larger than the effect of multiple scattering in the detector material and from resolution of the experimental measurements, as is the case for the STAR Detector within the measured kinematic range. 
The measured $\alpha$ distributions show broadening in hadronic Pb+Pb collisions with respect to UPCs. 
Figure~\ref{figure3} shows the $\alpha$ distributions from our calculations in Pb+Pb collisions at $\sqrt{s_{\rm_{NN}}} = $ 5.02 TeV for different centrality classes. 
The results are filtered with the fiducial cuts:  $p_{T\mu} > 4$ GeV/c, and $|\eta_{\mu}|<2.4$, and normalized to facilitate a direct comparison with experimental data from ATLAS. 
The measurements from ATLAS~\cite{PhysRevLett.121.212301} can be well described by the gEPA2 and QED calculations within uncertainties.

There have been proposals in the literature regarding possible final-state effects to explain the $P_\perp$ broadening. 
Two such proposals are that the broadening is due to deflection by the residual magnetic field trapped in an electrically conducting QGP~\cite{PhysRevLett.121.132301,ConductivityKharzeev:2009pj} or due to multiple Coulomb scattering in the hot and dense medium~\cite{PhysRevLett.121.212301,Klein:2018fmp}. 
All the proposed mechanisms including this study require extraordinarily strong electromagnetic fields, an interdisciplinary subject of intense interest across many scientific communities. 
There are a few assumptions and caveats in our calculation which deserve further studies: 
\begin{itemize}
    \item continuous charge distribution without point-like structure:\\
    It has been shown~\cite{Staig:2010by,Bzdak:2011yy} that the substructures of protons and quarks in nuclei and their fluctuations can significantly alter the electromagnetic field inside the nucleus at any given instant. This should result in an observable effect deserving further theoretical and experimental investigation. The effect is most prominent in central collisions where the ATLAS results have large uncertainties and where STAR currently lacks the necessary statistics for a measurement. 
    \item projectile and target nuclei maintain the same velocity vector before and after collision: \\
    The very first assumption in Eq.~\ref{equation1} is that both colliding nuclei maintain their velocities (a $\delta(k_{i}^{\nu}u_{i\nu})$ function) to simplify the calculation. In central collisions, where the photon flux are generated predominantly by the participant nucleons, charge stopping may be an important correction to the initial electromagnetic fields. 
    \item omission of higher order contribution and multiple pair production: \\
    We have ignored higher-order corrections in both the initial electromagnetic field~\cite{PhysRevC.70.031902} and Sudakov effect~\cite{Klein:2018fmp}, which should be quite small in the low $P_\bot$ and small $\alpha$ range. It has been pointed out that there may be significant multiple pair production in the same event~\cite{Hencken:2004td}, which may complicate the calculation and measurement. 
    \item final-state effects of magnetic field deflection and multiple Coulomb scattering: \\
    The STAR and ATLAS collaborations have demonstrated that it is possible to identify and measure the Breit-Wheeler process accompanying the creation of QGP. This opens new opportunity using this process as a probe of emerging QCD phenomena~\cite{ReportKharzeev:2015znc}.
\end{itemize}
In summary, we study the impact-parameter dependence of the Breit-Wheeler process in heavy-ion collisions within the framework of the external QED field and the approximations used to arrive at the Equivalent Photon Approximation, and with a full QED calculation based on two lowest-order Feynman diagrams. We further demonstrate that the $P_{\perp}$ spectrum from the STARLight model calculation used by the recent comparisons as a baseline results from averaging over the whole impact parameter space and is therefore by definition independent of impact parameter. Our model results can qualitatively describe both the $P_\perp$ broadening observed at RHIC as well as the acoplanarity broadening observed at the LHC. It provides a practical procedure for studying the Breit-Wheeler process with ultra-strong electromagnetic fields in a controllable fashion. This outcome indicates that the broadening originates predominantly from the initial electromagnetic field strength that varies significantly with impact parameter. There may be an additional small broadening due to final-state interaction. Determining precisely the magnitude of final-state effects, if they are present in the data, requires further verification of our QED calculation.  Future precision measurements from ultra-peripheral collisions would be useful for such verification of our QED calculations. 

The authors would like to thank Dr. Feng Yuan, Prof. Jinfeng Liao, Prof. Bowen Xiao and many members of STAR Collaboration for a stimulating discussion. We appreciate Prof. Jian Zhou and Ya-Jin Zhou for their  effort in double-checking our QED calculation and inspiring discussion on the $\cos{(4\phi)}$ modulation from polarized photon collisions~\cite{cos4phi}. 
This work was funded by the National Natural Science Foundation of China under Grant Nos. 11775213 and 11675168, the U.S. DOE Office of Science under contract No. DE-SC0012704 and DE-FG02-10ER41666, and MOST under Grant No. 2016YFE0104800.

\nocite{*}
\bibliographystyle{aipnum4-1}
\bibliography{EPACentrality}

\begin{thebibliography}{41}%
\makeatletter
\providecommand \@ifxundefined [1]{%
 \@ifx{#1\undefined}
}%
\providecommand \@ifnum [1]{%
 \ifnum #1\expandafter \@firstoftwo
 \else \expandafter \@secondoftwo
 \fi
}%
\providecommand \@ifx [1]{%
 \ifx #1\expandafter \@firstoftwo
 \else \expandafter \@secondoftwo
 \fi
}%
\providecommand \natexlab [1]{#1}%
\providecommand \enquote  [1]{``#1''}%
\providecommand \bibnamefont  [1]{#1}%
\providecommand \bibfnamefont [1]{#1}%
\providecommand \citenamefont [1]{#1}%
\providecommand \href@noop [0]{\@secondoftwo}%
\providecommand \href [0]{\begingroup \@sanitize@url \@href}%
\providecommand \@href[1]{\@@startlink{#1}\@@href}%
\providecommand \@@href[1]{\endgroup#1\@@endlink}%
\providecommand \@sanitize@url [0]{\catcode `\\12\catcode `\$12\catcode
  `\&12\catcode `\#12\catcode `\^12\catcode `\_12\catcode `\%12\relax}%
\providecommand \@@startlink[1]{}%
\providecommand \@@endlink[0]{}%
\providecommand \url  [0]{\begingroup\@sanitize@url \@url }%
\providecommand \@url [1]{\endgroup\@href {#1}{\urlprefix }}%
\providecommand \urlprefix  [0]{URL }%
\providecommand \Eprint [0]{\href }%
\providecommand \doibase [0]{http://dx.doi.org/}%
\providecommand \selectlanguage [0]{\@gobble}%
\providecommand \bibinfo  [0]{\@secondoftwo}%
\providecommand \bibfield  [0]{\@secondoftwo}%
\providecommand \translation [1]{[#1]}%
\providecommand \BibitemOpen [0]{}%
\providecommand \bibitemStop [0]{}%
\providecommand \bibitemNoStop [0]{.\EOS\space}%
\providecommand \EOS [0]{\spacefactor3000\relax}%
\providecommand \BibitemShut  [1]{\csname bibitem#1\endcsname}%
\let\auto@bib@innerbib\@empty
\bibitem [{\citenamefont {Breit}\ and\ \citenamefont
  {Wheeler}(1934)}]{Breit-wheeler1934zz}%
  \BibitemOpen
  \bibfield  {author} {\bibinfo {author} {\bibfnamefont {G.}~\bibnamefont
  {Breit}}\ and\ \bibinfo {author} {\bibfnamefont {J.~A.}\ \bibnamefont
  {Wheeler}},\ }\href {\doibase 10.1103/PhysRev.46.1087} {\bibfield  {journal}
  {\bibinfo  {journal} {Phys. Rev.}\ }\textbf {\bibinfo {volume} {46}},\
  \bibinfo {pages} {1087} (\bibinfo {year} {1934})}\BibitemShut {NoStop}%
\bibitem [{\citenamefont {Burke}\ \emph {et~al.}(1997)\citenamefont {Burke}
  \emph {et~al.}}]{SLACPhysRevLett.79.1626}%
  \BibitemOpen
  \bibfield  {author} {\bibinfo {author} {\bibfnamefont {D.~L.}\ \bibnamefont
  {Burke}} \emph {et~al.},\ }\href {\doibase 10.1103/PhysRevLett.79.1626}
  {\bibfield  {journal} {\bibinfo  {journal} {Phys. Rev. Lett.}\ }\textbf
  {\bibinfo {volume} {79}},\ \bibinfo {pages} {1626} (\bibinfo {year}
  {1997})}\BibitemShut {NoStop}%
\bibitem [{\citenamefont {Di~Piazza}\ \emph {et~al.}(2012)\citenamefont
  {Di~Piazza}, \citenamefont {M\"uller}, \citenamefont {Hatsagortsyan},\ and\
  \citenamefont {Keitel}}]{Piazza2012RevModPhys.84.1177}%
  \BibitemOpen
  \bibfield  {author} {\bibinfo {author} {\bibfnamefont {A.}~\bibnamefont
  {Di~Piazza}}, \bibinfo {author} {\bibfnamefont {C.}~\bibnamefont {M\"uller}},
  \bibinfo {author} {\bibfnamefont {K.~Z.}\ \bibnamefont {Hatsagortsyan}}, \
  and\ \bibinfo {author} {\bibfnamefont {C.~H.}\ \bibnamefont {Keitel}},\
  }\href {\doibase 10.1103/RevModPhys.84.1177} {\bibfield  {journal} {\bibinfo
  {journal} {Rev. Mod. Phys.}\ }\textbf {\bibinfo {volume} {84}},\ \bibinfo
  {pages} {1177} (\bibinfo {year} {2012})}\BibitemShut {NoStop}%
\bibitem [{\citenamefont {Pike}\ \emph {et~al.}(2014)\citenamefont {Pike},
  \citenamefont {Mackenroth}, \citenamefont {Hill},\ and\ \citenamefont
  {S.J.}}]{Pike2014NaturePhotonics}%
  \BibitemOpen
  \bibfield  {author} {\bibinfo {author} {\bibfnamefont {O.}~\bibnamefont
  {Pike}}, \bibinfo {author} {\bibfnamefont {F.}~\bibnamefont {Mackenroth}},
  \bibinfo {author} {\bibfnamefont {E.}~\bibnamefont {Hill}}, \ and\ \bibinfo
  {author} {\bibfnamefont {R.}~\bibnamefont {S.J.}},\ }\href {\doibase
  10.1038/nphoton.2014.95} {\bibfield  {journal} {\bibinfo  {journal} {Nature
  Photonics}\ }\textbf {\bibinfo {volume} {8}},\ \bibinfo {pages} {434}
  (\bibinfo {year} {2014})}\BibitemShut {NoStop}%
\bibitem [{\citenamefont {Baur}, \citenamefont {Hencken},\ and\ \citenamefont
  {Trautmann}(2007)}]{BAUR20071}%
  \BibitemOpen
  \bibfield  {author} {\bibinfo {author} {\bibfnamefont {G.}~\bibnamefont
  {Baur}}, \bibinfo {author} {\bibfnamefont {K.}~\bibnamefont {Hencken}}, \
  and\ \bibinfo {author} {\bibfnamefont {D.}~\bibnamefont {Trautmann}},\ }\href
  {\doibase https://doi.org/10.1016/j.physrep.2007.09.002} {\bibfield
  {journal} {\bibinfo  {journal} {Physics Reports}\ }\textbf {\bibinfo {volume}
  {453}},\ \bibinfo {pages} {1 } (\bibinfo {year} {2007})}\BibitemShut
  {NoStop}%
\bibitem [{\citenamefont {Krauss}, \citenamefont {Greiner},\ and\ \citenamefont
  {Soff}(1997)}]{KRAUSS1997503}%
  \BibitemOpen
  \bibfield  {author} {\bibinfo {author} {\bibfnamefont {F.}~\bibnamefont
  {Krauss}}, \bibinfo {author} {\bibfnamefont {M.}~\bibnamefont {Greiner}}, \
  and\ \bibinfo {author} {\bibfnamefont {G.}~\bibnamefont {Soff}},\ }\href
  {\doibase https://doi.org/10.1016/S0146-6410(97)00049-5} {\bibfield
  {journal} {\bibinfo  {journal} {Progress in Particle and Nuclear Physics}\
  }\textbf {\bibinfo {volume} {39}},\ \bibinfo {pages} {503 } (\bibinfo {year}
  {1997})}\BibitemShut {NoStop}%
\bibitem [{\citenamefont {Kharzeev}, \citenamefont {McLerran},\ and\
  \citenamefont {Warringa}(2008)}]{BFieldKharzeev:2007jp}%
  \BibitemOpen
  \bibfield  {author} {\bibinfo {author} {\bibfnamefont {D.~E.}\ \bibnamefont
  {Kharzeev}}, \bibinfo {author} {\bibfnamefont {L.~D.}\ \bibnamefont
  {McLerran}}, \ and\ \bibinfo {author} {\bibfnamefont {H.~J.}\ \bibnamefont
  {Warringa}},\ }\href {\doibase 10.1016/j.nuclphysa.2008.02.298} {\bibfield
  {journal} {\bibinfo  {journal} {Nucl. Phys.}\ }\textbf {\bibinfo {volume}
  {A803}},\ \bibinfo {pages} {227} (\bibinfo {year} {2008})},\ \Eprint
  {http://arxiv.org/abs/0711.0950} {arXiv:0711.0950 [hep-ph]} \BibitemShut
  {NoStop}%
\bibitem [{\citenamefont {Kharzeev}\ \emph {et~al.}(2016)\citenamefont
  {Kharzeev}, \citenamefont {Liao}, \citenamefont {Voloshin},\ and\
  \citenamefont {Wang}}]{ReportKharzeev:2015znc}%
  \BibitemOpen
  \bibfield  {author} {\bibinfo {author} {\bibfnamefont {D.~E.}\ \bibnamefont
  {Kharzeev}}, \bibinfo {author} {\bibfnamefont {J.}~\bibnamefont {Liao}},
  \bibinfo {author} {\bibfnamefont {S.~A.}\ \bibnamefont {Voloshin}}, \ and\
  \bibinfo {author} {\bibfnamefont {G.}~\bibnamefont {Wang}},\ }\href {\doibase
  10.1016/j.ppnp.2016.01.001} {\bibfield  {journal} {\bibinfo  {journal} {Prog.
  Part. Nucl. Phys.}\ }\textbf {\bibinfo {volume} {88}},\ \bibinfo {pages} {1}
  (\bibinfo {year} {2016})},\ \Eprint {http://arxiv.org/abs/1511.04050}
  {arXiv:1511.04050 [hep-ph]} \BibitemShut {NoStop}%
\bibitem [{\citenamefont {Vidovi\ifmmode~\acute{c}\else \'{c}\fi{}}\ \emph
  {et~al.}(1993)\citenamefont {Vidovi\ifmmode~\acute{c}\else \'{c}\fi{}} \emph
  {et~al.}}]{PhysRevC.47.2308}%
  \BibitemOpen
  \bibfield  {author} {\bibinfo {author} {\bibfnamefont {M.}~\bibnamefont
  {Vidovi\ifmmode~\acute{c}\else \'{c}\fi{}}} \emph {et~al.},\ }\href {\doibase
  10.1103/PhysRevC.47.2308} {\bibfield  {journal} {\bibinfo  {journal} {Phys.
  Rev. C}\ }\textbf {\bibinfo {volume} {47}},\ \bibinfo {pages} {2308}
  (\bibinfo {year} {1993})}\BibitemShut {NoStop}%
\bibitem [{\citenamefont {Adams}\ \emph {et~al.}(2004)\citenamefont {Adams}
  \emph {et~al.}}]{PhysRevC.70.031902}%
  \BibitemOpen
  \bibfield  {author} {\bibinfo {author} {\bibfnamefont {J.}~\bibnamefont
  {Adams}} \emph {et~al.} (\bibinfo {collaboration} {STAR Collaboration}),\
  }\href {\doibase 10.1103/PhysRevC.70.031902} {\bibfield  {journal} {\bibinfo
  {journal} {Phys. Rev. C}\ }\textbf {\bibinfo {volume} {70}},\ \bibinfo
  {pages} {031902} (\bibinfo {year} {2004})}\BibitemShut {NoStop}%
\bibitem [{\citenamefont {Dyndal}(2017)}]{DYNDAL2017281}%
  \BibitemOpen
  \bibfield  {author} {\bibinfo {author} {\bibfnamefont {M.}~\bibnamefont
  {Dyndal}} (\bibinfo {collaboration} {ATLAS Collaboration}),\ }\href {\doibase
  https://doi.org/10.1016/j.nuclphysa.2017.04.043} {\bibfield  {journal}
  {\bibinfo  {journal} {Nuclear Physics A}\ }\textbf {\bibinfo {volume}
  {967}},\ \bibinfo {pages} {281 } (\bibinfo {year} {2017})}\BibitemShut
  {NoStop}%
\bibitem [{\citenamefont {Abbas}\ \emph {et~al.}(2013)\citenamefont {Abbas}
  \emph {et~al.}}]{Abbas2013}%
  \BibitemOpen
  \bibfield  {author} {\bibinfo {author} {\bibfnamefont {E.}~\bibnamefont
  {Abbas}} \emph {et~al.} (\bibinfo {collaboration} {ALICE Collaboration}),\
  }\href {\doibase 10.1140/epjc/s10052-013-2617-1} {\bibfield  {journal}
  {\bibinfo  {journal} {The European Physical Journal C}\ }\textbf {\bibinfo
  {volume} {73}},\ \bibinfo {pages} {2617} (\bibinfo {year}
  {2013})}\BibitemShut {NoStop}%
\bibitem [{\citenamefont {Khachatryan}\ \emph {et~al.}(2017)\citenamefont
  {Khachatryan} \emph {et~al.}}]{2017489}%
  \BibitemOpen
  \bibfield  {author} {\bibinfo {author} {\bibfnamefont {V.}~\bibnamefont
  {Khachatryan}} \emph {et~al.} (\bibinfo {collaboration} {CMS
  Collaboration}),\ }\href {\doibase
  https://doi.org/10.1016/j.physletb.2017.07.001} {\bibfield  {journal}
  {\bibinfo  {journal} {Physics Letters B}\ }\textbf {\bibinfo {volume}
  {772}},\ \bibinfo {pages} {489 } (\bibinfo {year} {2017})}\BibitemShut
  {NoStop}%
\bibitem [{\citenamefont {Baur}\ and\ \citenamefont
  {Filho}(1990)}]{BAUR1990786}%
  \BibitemOpen
  \bibfield  {author} {\bibinfo {author} {\bibfnamefont {G.}~\bibnamefont
  {Baur}}\ and\ \bibinfo {author} {\bibfnamefont {L.}~\bibnamefont {Filho}},\
  }\href {\doibase https://doi.org/10.1016/0375-9474(90)90191-N} {\bibfield
  {journal} {\bibinfo  {journal} {Nuclear Physics A}\ }\textbf {\bibinfo
  {volume} {518}},\ \bibinfo {pages} {786 } (\bibinfo {year}
  {1990})}\BibitemShut {NoStop}%
\bibitem [{\citenamefont {Klein}\ \emph {et~al.}(2017)\citenamefont {Klein}
  \emph {et~al.}}]{Klein:2016yzr}%
  \BibitemOpen
  \bibfield  {author} {\bibinfo {author} {\bibfnamefont {S.~R.}\ \bibnamefont
  {Klein}} \emph {et~al.},\ }\href {\doibase 10.1016/j.cpc.2016.10.016}
  {\bibfield  {journal} {\bibinfo  {journal} {Comput. Phys. Commun.}\ }\textbf
  {\bibinfo {volume} {212}},\ \bibinfo {pages} {258} (\bibinfo {year}
  {2017})}\BibitemShut {NoStop}%
\bibitem [{\citenamefont {Baltz}\ \emph {et~al.}(2009)\citenamefont {Baltz}
  \emph {et~al.}}]{PhysRevC.80.044902}%
  \BibitemOpen
  \bibfield  {author} {\bibinfo {author} {\bibfnamefont {A.~J.}\ \bibnamefont
  {Baltz}} \emph {et~al.},\ }\href {\doibase 10.1103/PhysRevC.80.044902}
  {\bibfield  {journal} {\bibinfo  {journal} {Phys. Rev. C}\ }\textbf {\bibinfo
  {volume} {80}},\ \bibinfo {pages} {044902} (\bibinfo {year}
  {2009})}\BibitemShut {NoStop}%
\bibitem [{\citenamefont {Hencken}, \citenamefont {Trautmann},\ and\
  \citenamefont {Baur}(1995{\natexlab{a}})}]{Hencken:1995me}%
  \BibitemOpen
  \bibfield  {author} {\bibinfo {author} {\bibfnamefont {K.}~\bibnamefont
  {Hencken}}, \bibinfo {author} {\bibfnamefont {D.}~\bibnamefont {Trautmann}},
  \ and\ \bibinfo {author} {\bibfnamefont {G.}~\bibnamefont {Baur}},\ }\href
  {\doibase 10.1007/BF01620724} {\bibfield  {journal} {\bibinfo  {journal} {Z.
  Phys.}\ }\textbf {\bibinfo {volume} {C68}},\ \bibinfo {pages} {473} (\bibinfo
  {year} {1995}{\natexlab{a}})}\BibitemShut {NoStop}%
\bibitem [{\citenamefont {Grabiak}\ \emph {et~al.}(1989)\citenamefont {Grabiak}
  \emph {et~al.}}]{0954-3899-15-3-001}%
  \BibitemOpen
  \bibfield  {author} {\bibinfo {author} {\bibfnamefont {M.}~\bibnamefont
  {Grabiak}} \emph {et~al.},\ }\href
  {http://stacks.iop.org/0954-3899/15/i=3/a=001} {\bibfield  {journal}
  {\bibinfo  {journal} {Journal of Physics G: Nuclear and Particle Physics}\
  }\textbf {\bibinfo {volume} {15}},\ \bibinfo {pages} {L25} (\bibinfo {year}
  {1989})}\BibitemShut {NoStop}%
\bibitem [{\citenamefont {Alscher}\ \emph
  {et~al.}(1997{\natexlab{a}})\citenamefont {Alscher} \emph
  {et~al.}}]{Alscher:1996mja}%
  \BibitemOpen
  \bibfield  {author} {\bibinfo {author} {\bibfnamefont {A.}~\bibnamefont
  {Alscher}} \emph {et~al.},\ }\href {\doibase 10.1103/PhysRevA.55.396}
  {\bibfield  {journal} {\bibinfo  {journal} {Phys. Rev.}\ }\textbf {\bibinfo
  {volume} {A55}},\ \bibinfo {pages} {396} (\bibinfo {year}
  {1997}{\natexlab{a}})}\BibitemShut {NoStop}%
\bibitem [{\citenamefont {Aaboud}\ \emph {et~al.}(2017)\citenamefont {Aaboud}
  \emph {et~al.}}]{ATLASNatureAaboud:2017bwk}%
  \BibitemOpen
  \bibfield  {author} {\bibinfo {author} {\bibfnamefont {M.}~\bibnamefont
  {Aaboud}} \emph {et~al.} (\bibinfo {collaboration} {ATLAS}),\ }\href
  {\doibase 10.1038/nphys4208} {\bibfield  {journal} {\bibinfo  {journal}
  {Nature Phys.}\ }\textbf {\bibinfo {volume} {13}},\ \bibinfo {pages} {852}
  (\bibinfo {year} {2017})},\ \Eprint {http://arxiv.org/abs/1702.01625}
  {arXiv:1702.01625 [hep-ex]} \BibitemShut {NoStop}%
\bibitem [{\citenamefont {Adam}\ \emph {et~al.}(2016)\citenamefont {Adam} \emph
  {et~al.}}]{LOW_ALICE}%
  \BibitemOpen
  \bibfield  {author} {\bibinfo {author} {\bibfnamefont {J.}~\bibnamefont
  {Adam}} \emph {et~al.} (\bibinfo {collaboration} {ALICE Collaboration}),\
  }\href {<Go to ISI>://WOS:000377018100005
  http://journals.aps.org/prl/pdf/10.1103/PhysRevLett.116.222301} {\bibfield
  {journal} {\bibinfo  {journal} {Physical Review Letters}\ }\textbf {\bibinfo
  {volume} {116}},\ \bibinfo {pages} {222301} (\bibinfo {year}
  {2016})}\BibitemShut {NoStop}%
\bibitem [{\citenamefont {Zha}(2017)}]{1742-6596-779-1-012039}%
  \BibitemOpen
  \bibfield  {author} {\bibinfo {author} {\bibfnamefont {W.}~\bibnamefont
  {Zha}} (\bibinfo {collaboration} {STAR Collaboration}),\ }\href
  {http://stacks.iop.org/1742-6596/779/i=1/a=012039} {\bibfield  {journal}
  {\bibinfo  {journal} {Journal of Physics: Conference Series}\ }\textbf
  {\bibinfo {volume} {779}},\ \bibinfo {pages} {012039} (\bibinfo {year}
  {2017})}\BibitemShut {NoStop}%
\bibitem [{\citenamefont {Yang}(2016)}]{syang2016thesis}%
  \BibitemOpen
  \bibfield  {author} {\bibinfo {author} {\bibfnamefont {S.}~\bibnamefont
  {Yang}},\ }\emph {\bibinfo {title} {Dielectron production in U+U collisions
  at $\sqrt{s}$=193 GeV at RHIC}},\ \href@noop {} {Ph.D. thesis},\ \bibinfo
  {school} {University of Science and Technology of China} (\bibinfo {year}
  {2016})\BibitemShut {NoStop}%
\bibitem [{\citenamefont {Adam}\ \emph {et~al.}(2018)\citenamefont {Adam} \emph
  {et~al.}}]{PhysRevLett.121.132301}%
  \BibitemOpen
  \bibfield  {author} {\bibinfo {author} {\bibfnamefont {J.}~\bibnamefont
  {Adam}} \emph {et~al.} (\bibinfo {collaboration} {STAR Collaboration}),\
  }\href {\doibase 10.1103/PhysRevLett.121.132301} {\bibfield  {journal}
  {\bibinfo  {journal} {Phys. Rev. Lett.}\ }\textbf {\bibinfo {volume} {121}},\
  \bibinfo {pages} {132301} (\bibinfo {year} {2018})}\BibitemShut {NoStop}%
\bibitem [{\citenamefont {Aaboud}\ \emph {et~al.}(2018)\citenamefont {Aaboud}
  \emph {et~al.}}]{PhysRevLett.121.212301}%
  \BibitemOpen
  \bibfield  {author} {\bibinfo {author} {\bibfnamefont {M.}~\bibnamefont
  {Aaboud}} \emph {et~al.} (\bibinfo {collaboration} {ATLAS Collaboration}),\
  }\href {\doibase 10.1103/PhysRevLett.121.212301} {\bibfield  {journal}
  {\bibinfo  {journal} {Phys. Rev. Lett.}\ }\textbf {\bibinfo {volume} {121}},\
  \bibinfo {pages} {212301} (\bibinfo {year} {2018})}\BibitemShut {NoStop}%
\bibitem [{\citenamefont {K\l{}usek-Gawenda}\ and\ \citenamefont
  {Szczurek}(2016)}]{PhysRevC.93.044912}%
  \BibitemOpen
  \bibfield  {author} {\bibinfo {author} {\bibfnamefont {M.}~\bibnamefont
  {K\l{}usek-Gawenda}}\ and\ \bibinfo {author} {\bibfnamefont {A.}~\bibnamefont
  {Szczurek}},\ }\href {\doibase 10.1103/PhysRevC.93.044912} {\bibfield
  {journal} {\bibinfo  {journal} {Phys. Rev. C}\ }\textbf {\bibinfo {volume}
  {93}},\ \bibinfo {pages} {044912} (\bibinfo {year} {2016})}\BibitemShut
  {NoStop}%
\bibitem [{\citenamefont {Zha}\ \emph {et~al.}(2018{\natexlab{a}})\citenamefont
  {Zha} \emph {et~al.}}]{PhysRevC.97.044910}%
  \BibitemOpen
  \bibfield  {author} {\bibinfo {author} {\bibfnamefont {W.}~\bibnamefont
  {Zha}} \emph {et~al.},\ }\href {\doibase 10.1103/PhysRevC.97.044910}
  {\bibfield  {journal} {\bibinfo  {journal} {Phys. Rev. C}\ }\textbf {\bibinfo
  {volume} {97}},\ \bibinfo {pages} {044910} (\bibinfo {year}
  {2018}{\natexlab{a}})}\BibitemShut {NoStop}%
\bibitem [{\citenamefont {Zha}\ \emph {et~al.}(2018{\natexlab{b}})\citenamefont
  {Zha} \emph {et~al.}}]{ZHA2018182}%
  \BibitemOpen
  \bibfield  {author} {\bibinfo {author} {\bibfnamefont {W.}~\bibnamefont
  {Zha}} \emph {et~al.},\ }\href {\doibase
  https://doi.org/10.1016/j.physletb.2018.04.006} {\bibfield  {journal}
  {\bibinfo  {journal} {Physics Letters B}\ }\textbf {\bibinfo {volume}
  {781}},\ \bibinfo {pages} {182 } (\bibinfo {year}
  {2018}{\natexlab{b}})}\BibitemShut {NoStop}%
\bibitem [{\citenamefont {Klein}(2018)}]{PhysRevC.97.054903}%
  \BibitemOpen
  \bibfield  {author} {\bibinfo {author} {\bibfnamefont {S.~R.}\ \bibnamefont
  {Klein}},\ }\href {\doibase 10.1103/PhysRevC.97.054903} {\bibfield  {journal}
  {\bibinfo  {journal} {Phys. Rev. C}\ }\textbf {\bibinfo {volume} {97}},\
  \bibinfo {pages} {054903} (\bibinfo {year} {2018})}\BibitemShut {NoStop}%
\bibitem [{\citenamefont {Hencken}, \citenamefont {Trautmann},\ and\
  \citenamefont {Baur}(1995{\natexlab{b}})}]{PhysRevA.51.1874}%
  \BibitemOpen
  \bibfield  {author} {\bibinfo {author} {\bibfnamefont {K.}~\bibnamefont
  {Hencken}}, \bibinfo {author} {\bibfnamefont {D.}~\bibnamefont {Trautmann}},
  \ and\ \bibinfo {author} {\bibfnamefont {G.}~\bibnamefont {Baur}},\ }\href
  {\doibase 10.1103/PhysRevA.51.1874} {\bibfield  {journal} {\bibinfo
  {journal} {Phys. Rev. A}\ }\textbf {\bibinfo {volume} {51}},\ \bibinfo
  {pages} {1874} (\bibinfo {year} {1995}{\natexlab{b}})}\BibitemShut {NoStop}%
\bibitem [{\citenamefont {Alscher}\ \emph
  {et~al.}(1997{\natexlab{b}})\citenamefont {Alscher}, \citenamefont {Hencken},
  \citenamefont {Trautmann},\ and\ \citenamefont {Baur}}]{PhysRevA.55.396}%
  \BibitemOpen
  \bibfield  {author} {\bibinfo {author} {\bibfnamefont {A.}~\bibnamefont
  {Alscher}}, \bibinfo {author} {\bibfnamefont {K.}~\bibnamefont {Hencken}},
  \bibinfo {author} {\bibfnamefont {D.}~\bibnamefont {Trautmann}}, \ and\
  \bibinfo {author} {\bibfnamefont {G.}~\bibnamefont {Baur}},\ }\href {\doibase
  10.1103/PhysRevA.55.396} {\bibfield  {journal} {\bibinfo  {journal} {Phys.
  Rev. A}\ }\textbf {\bibinfo {volume} {55}},\ \bibinfo {pages} {396} (\bibinfo
  {year} {1997}{\natexlab{b}})}\BibitemShut {NoStop}%
\bibitem [{\citenamefont {Barrett}\ and\ \citenamefont
  {Jackson}(1977)}]{0031-9112-29-7-028}%
  \BibitemOpen
  \bibfield  {author} {\bibinfo {author} {\bibfnamefont {R.~C.}\ \bibnamefont
  {Barrett}}\ and\ \bibinfo {author} {\bibfnamefont {D.~F.}\ \bibnamefont
  {Jackson}},\ }\href@noop {} {\emph {\bibinfo {title} {Nuclear Sizes and
  Structure}}}\ (\bibinfo  {publisher} {Oxford University Press},\ \bibinfo
  {year} {1977})\BibitemShut {NoStop}%
\bibitem [{\citenamefont {Klein}\ \emph {et~al.}(2019)\citenamefont {Klein},
  \citenamefont {Mueller}, \citenamefont {Xiao},\ and\ \citenamefont
  {Yuan}}]{Klein:2018fmp}%
  \BibitemOpen
  \bibfield  {author} {\bibinfo {author} {\bibfnamefont {S.}~\bibnamefont
  {Klein}}, \bibinfo {author} {\bibfnamefont {A.}~\bibnamefont {Mueller}},
  \bibinfo {author} {\bibfnamefont {B.-W.}\ \bibnamefont {Xiao}}, \ and\
  \bibinfo {author} {\bibfnamefont {F.}~\bibnamefont {Yuan}},\ }\href {\doibase
  10.1103/PhysRevLett.122.132301} {\bibfield  {journal} {\bibinfo  {journal}
  {Phys. Rev. Lett.}\ }\textbf {\bibinfo {volume} {122}},\ \bibinfo {pages}
  {132301} (\bibinfo {year} {2019})}\BibitemShut {NoStop}%
\bibitem [{\citenamefont {Shtabovenko}, \citenamefont {Mertig},\ and\
  \citenamefont {Orellana}(2016)}]{SHTABOVENKO2016432}%
  \BibitemOpen
  \bibfield  {author} {\bibinfo {author} {\bibfnamefont {V.}~\bibnamefont
  {Shtabovenko}}, \bibinfo {author} {\bibfnamefont {R.}~\bibnamefont {Mertig}},
  \ and\ \bibinfo {author} {\bibfnamefont {F.}~\bibnamefont {Orellana}},\
  }\href {\doibase https://doi.org/10.1016/j.cpc.2016.06.008} {\bibfield
  {journal} {\bibinfo  {journal} {Computer Physics Communications}\ }\textbf
  {\bibinfo {volume} {207}},\ \bibinfo {pages} {432 } (\bibinfo {year}
  {2016})}\BibitemShut {NoStop}%
\bibitem [{\citenamefont {Lepage}(1978)}]{PETERLEPAGE1978192}%
  \BibitemOpen
  \bibfield  {author} {\bibinfo {author} {\bibfnamefont {G.~P.}\ \bibnamefont
  {Lepage}},\ }\href {\doibase https://doi.org/10.1016/0021-9991(78)90004-9}
  {\bibfield  {journal} {\bibinfo  {journal} {Journal of Computational
  Physics}\ }\textbf {\bibinfo {volume} {27}},\ \bibinfo {pages} {192 }
  (\bibinfo {year} {1978})}\BibitemShut {NoStop}%
\bibitem [{\citenamefont {Hencken}, \citenamefont {Baur},\ and\ \citenamefont
  {Trautmann}(2004)}]{Hencken:2004td}%
  \BibitemOpen
  \bibfield  {author} {\bibinfo {author} {\bibfnamefont {K.}~\bibnamefont
  {Hencken}}, \bibinfo {author} {\bibfnamefont {G.}~\bibnamefont {Baur}}, \
  and\ \bibinfo {author} {\bibfnamefont {D.}~\bibnamefont {Trautmann}},\ }\href
  {\doibase 10.1103/PhysRevC.69.054902} {\bibfield  {journal} {\bibinfo
  {journal} {Phys. Rev.}\ }\textbf {\bibinfo {volume} {C69}},\ \bibinfo {pages}
  {054902} (\bibinfo {year} {2004})},\ \Eprint
  {http://arxiv.org/abs/nucl-th/0402061} {arXiv:nucl-th/0402061 [nucl-th]}
  \BibitemShut {NoStop}%
\bibitem [{\citenamefont {Kharzeev}\ and\ \citenamefont
  {Warringa}(2009)}]{ConductivityKharzeev:2009pj}%
  \BibitemOpen
  \bibfield  {author} {\bibinfo {author} {\bibfnamefont {D.~E.}\ \bibnamefont
  {Kharzeev}}\ and\ \bibinfo {author} {\bibfnamefont {H.~J.}\ \bibnamefont
  {Warringa}},\ }\href {\doibase 10.1103/PhysRevD.80.034028} {\bibfield
  {journal} {\bibinfo  {journal} {Phys. Rev.}\ }\textbf {\bibinfo {volume}
  {D80}},\ \bibinfo {pages} {034028} (\bibinfo {year} {2009})},\ \Eprint
  {http://arxiv.org/abs/0907.5007} {arXiv:0907.5007 [hep-ph]} \BibitemShut
  {NoStop}%
\bibitem [{\citenamefont {Staig}\ and\ \citenamefont
  {Shuryak}(2010)}]{Staig:2010by}%
  \BibitemOpen
  \bibfield  {author} {\bibinfo {author} {\bibfnamefont {P.}~\bibnamefont
  {Staig}}\ and\ \bibinfo {author} {\bibfnamefont {E.}~\bibnamefont
  {Shuryak}},\ }\href@noop {} {\  (\bibinfo {year} {2010})},\ \Eprint
  {http://arxiv.org/abs/1005.3531} {arXiv:1005.3531 [nucl-th]} \BibitemShut
  {NoStop}%
\bibitem [{\citenamefont {Bzdak}\ and\ \citenamefont
  {Skokov}(2012)}]{Bzdak:2011yy}%
  \BibitemOpen
  \bibfield  {author} {\bibinfo {author} {\bibfnamefont {A.}~\bibnamefont
  {Bzdak}}\ and\ \bibinfo {author} {\bibfnamefont {V.}~\bibnamefont {Skokov}},\
  }\href {\doibase 10.1016/j.physletb.2012.02.065} {\bibfield  {journal}
  {\bibinfo  {journal} {Phys. Lett.}\ }\textbf {\bibinfo {volume} {B710}},\
  \bibinfo {pages} {171} (\bibinfo {year} {2012})},\ \Eprint
  {http://arxiv.org/abs/1111.1949} {arXiv:1111.1949 [hep-ph]} \BibitemShut
  {NoStop}%
\bibitem [{\citenamefont {Li}, \citenamefont {Zhou},\ and\ \citenamefont
  {Zhou}(2019)}]{cos4phi}%
  \BibitemOpen
  \bibfield  {author} {\bibinfo {author} {\bibfnamefont {C.}~\bibnamefont
  {Li}}, \bibinfo {author} {\bibfnamefont {J.}~\bibnamefont {Zhou}}, \ and\
  \bibinfo {author} {\bibfnamefont {Y.-J.}\ \bibnamefont {Zhou}},\ }\href@noop
  {} {\bibfield  {journal} {\bibinfo  {journal} {arXiv:1903.10084}\ } (\bibinfo
  {year} {2019})},\ \Eprint {http://arxiv.org/abs/1903.10084} {arXiv:1903.10084
  [hep-ph]} \BibitemShut {NoStop}%
\bibitem [{\citenamefont {Ackermann}\ \emph {et~al.}(2001)\citenamefont
  {Ackermann} \emph {et~al.}}]{STARv2Ackermann:2000tr}%
  \BibitemOpen
  \bibfield  {author} {\bibinfo {author} {\bibfnamefont {K.~H.}\ \bibnamefont
  {Ackermann}} \emph {et~al.} (\bibinfo {collaboration} {STAR}),\ }\href
  {\doibase 10.1103/PhysRevLett.86.402} {\bibfield  {journal} {\bibinfo
  {journal} {Phys. Rev. Lett.}\ }\textbf {\bibinfo {volume} {86}},\ \bibinfo
  {pages} {402} (\bibinfo {year} {2001})},\ \Eprint
  {http://arxiv.org/abs/nucl-ex/0009011} {arXiv:nucl-ex/0009011 [nucl-ex]}
  \BibitemShut {NoStop}%
\end{thebibliography}%
\end{document}